\documentclass{emulateapj}

\usepackage{natbib} 
\usepackage{multirow} 
\usepackage[caption=false]{subfig} 
\usepackage{enumerate} 
\usepackage[T1]{fontenc} 
\usepackage{amsmath} 

\def\reff@jnl#1{{\rm#1\/}}

\def\aj{\reff@jnl{AJ}}                  
\def\araa{\reff@jnl{ARA\&A}}            
\def\apj{\reff@jnl{ApJ}}                
\def\apjl{\reff@jnl{ApJ}}               
\def\apjs{\reff@jnl{ApJS}}              
\def\ao{\reff@jnl{Appl.Optics}}         
\def\apss{\reff@jnl{Ap\&SS}}            
\def\aap{\reff@jnl{A\&A}}               
\def\aapr{\reff@jnl{A\&A~Rev.}}         
\def\aaps{\reff@jnl{A\&AS}}             
\def\azh{\reff@jnl{AZh}}                        
\def\baas{\reff@jnl{BAAS}}              
\def\jrasc{\reff@jnl{JRASC}}            
\def\memras{\reff@jnl{MmRAS}}           
\def\mnras{\reff@jnl{MNRAS}}            
\def\pra{\reff@jnl{Phys.Rev.A}}         
\def\prb{\reff@jnl{Phys.Rev.B}}         
\def\prc{\reff@jnl{Phys.Rev.C}}         
\def\prd{\reff@jnl{Phys.Rev.D}}         
\def\prl{\reff@jnl{Phys.Rev.Lett}}      
\def\pasp{\reff@jnl{PASP}}              
\def\pasj{\reff@jnl{PASJ}}              
\def\qjras{\reff@jnl{QJRAS}}            
\def\skytel{\reff@jnl{S\&T}}            
\def\solphys{\reff@jnl{Solar~Phys.}}    
\def\sovast{\reff@jnl{Soviet~Ast.}}     
\def\ssr{\reff@jnl{Space~Sci.Rev.}}     
\def\zap{\reff@jnl{ZAp}}                        
\def\nat{\reff@jnl{Nature}}             

\begin{document}


\title{Constraints on free-free emission from Anomalous Microwave Emission Sources in the Perseus Molecular Cloud}

\author{C.~T.~Tibbs\altaffilmark{1,2}, R.~Paladini\altaffilmark{3}, C.~Dickinson\altaffilmark{2}, B.~S.~Mason\altaffilmark{4}, S.~Casassus\altaffilmark{5}, K.~Cleary\altaffilmark{6}, R.~D.~Davies\altaffilmark{2}, R.~J.~Davis\altaffilmark{2}, R.~A.~Watson\altaffilmark{2}}
\email{ctibbs@ipac.caltech.edu}

\altaffiltext{1}{\textit{Spitzer} Science Center, California Institute of Technology, Pasadena, CA 91125, USA}
\altaffiltext{2}{Jodrell Bank Centre for Astrophysics, School of Physics and Astronomy, The University of Manchester, Manchester, M13 9PL, UK}
\altaffiltext{3}{NASA \textit{Herschel} Science Center, California Institute of Technology, Pasadena, CA 91125, USA}
\altaffiltext{4}{National Radio Astronomy Observatory, 520 Edgemont Road, Charlottesville, VA 22903, USA}
\altaffiltext{5}{Departamento de Astronom{\'{\i}}a, Universidad de Chile, Casilla 36-D, Santiago, Chile}
\altaffiltext{6}{Cahill Center for Astronomy and Astrophysics, California Institute of Technology, Pasadena, CA 91125, USA}

\shorttitle{GBT Observations of the Perseus Cloud}
\shortauthors{Tibbs et al.}


\begin{abstract}

We present observations performed with the Green Bank Telescope at 1.4 and 5~GHz of three strips coincident with the anomalous microwave emission features previously identified in the Perseus molecular cloud at 33~GHz with the Very Small Array. With these observations we determine the level of the low frequency~($\sim$~1~--~5~GHz) emission. We do not detect any significant extended emission in these regions and we compute conservative 3$\sigma$ upper limits on the fraction of free-free emission at 33~GHz of 27~\%, 12~\%, and 18~\% for the three strips, indicating that the level of the emission at 1.4 and 5~GHz cannot account for the emission observed at 33~GHz. Additionally, we find that the low frequency emission is not spatially correlated with the emission observed at 33~GHz. These results indicate that the emission observed in the Perseus molecular cloud at 33~GHz, is indeed in excess over the low frequency emission, hence confirming its anomalous nature.

\end{abstract}




\section{Introduction}
\label{sec:intro}

In recent years, anomalous microwave emission~(AME) has been established as a new Galactic emission mechanism. Identified as an excess of emission between~$\sim$~10~--~100~GHz, AME detections require observations at frequencies both above and below this range to determine the level of the other Galactic emission mechanisms. At frequencies below 10~GHz, emission from the interactions between the free electrons and ions in ionized gas and emission from the acceleration of relativistic electrons in the Galactic magnetic fields dominate, while emission above 100~GHz is due to the thermal emission from big, interstellar dust grains in thermal equilibrium with the exciting radiation field.

Although AME has been found in numerous Galactic objects~\citep[e.g.][]{Casassus:08, Ami:09, Scaife:09, Dickinson:10, Tibbs:10, Planck_Dickinson:11, Tibbs:12} and in diffuse environments at high Galactic latitudes~\citep[e.g.][]{Ghosh:12, Peel:12}, there is still much to learn about this enigmatic emission mechanism. AME has been found to be highly spatially correlated with the dust emitting at infrared~(IR) wavelengths, indicating a direct association with interstellar dust grains, and at present there are two viable explanations for the AME: 1) electric dipole emission due to the rotation of small dust grains characterized by an electric dipole moment~\citep{DaL:98, Ali-Haimoud:09, Hoang:10, Ysard:10, Hoang:11, Silsbee:11}; and 2) magnetic dipole emission due to fluctuations in the magnetism of dust grains containing magnetic materials~\citep{DaL:99, Draine:13}. Of these two emission mechanisms, electric dipole emission from spinning dust grains, commonly referred to as spinning dust emission, is the explanation currently favored by observations.

In this work we focus on the Perseus molecular cloud, which has previously been studied in detail and found to be a source of AME~\citep{Watson:05, Tibbs:10, Planck_Dickinson:11, Tibbs:13}. AME was first detected in this cloud by~\citet{Watson:05}, who combined observations performed with the COSMOSOMAS Experiment with data from low frequency radio surveys, \textit{WMAP} and DIRBE, to produce a complete spectrum for the cloud from the radio to the IR on angular scales of~$\sim$~1~degree. This spectrum exhibited a clear excess of emission between~$\sim$~10~--~60~GHz, that was well fitted by spinning dust models. Follow-up observations of this region performed at 33~GHz with the Very Small Array~(VSA) interferometer by~\citet{Tibbs:10} found excess emission in five features on angular scales of~$\sim~$10~--~40~arcmin. The authors found that the total emission observed with the VSA in these five features accounted for only~$\sim$~10~\% of the emission detected on degree angular scales by~\citet{Watson:05}. In their analysis,~\citet{Tibbs:10} used the GB6 all-sky survey at 4.85~GHz~\citep{Condon:89} to constrain the low frequency emission in the five features as these were the only suitable observations available. However, as pointed out in that analysis, the GB6 data have been filtered to remove emission on angular scales greater than~$\sim$~20~arcmin. Therefore, the VSA data had to be filtered to match the range of angular scales to those which the GB6 observations were sensitive, before the level of the low frequency emission was determined. Here we present new observations of the five AME features with the Robert C. Byrd Green Bank Telescope~(GBT) at 1.4 and 5~GHz. These new observations allow us to directly investigate the level of emission at 1.4 and 5~GHz on the full range of angular scales observed with the VSA. With these observations, we investigate the spatial structure of the low frequency~(1.4 and 5~GHz) emission and how it compares to the five AME features observed with the VSA at 33~GHz.

The layout of this paper is as follows. In Section~\ref{sec:obs} we describe the GBT observations and data reduction. In Section~\ref{sec:discuss} we investigate the level of the low frequency emission with respect to the 33~GHz emission, and in Section~\ref{sec:con} we present our conclusions.


\section{GBT Observations}
\label{sec:obs}

\begin{figure}
\begin{center}
\includegraphics[angle=0,scale=0.48, viewport=10 0 600 600]{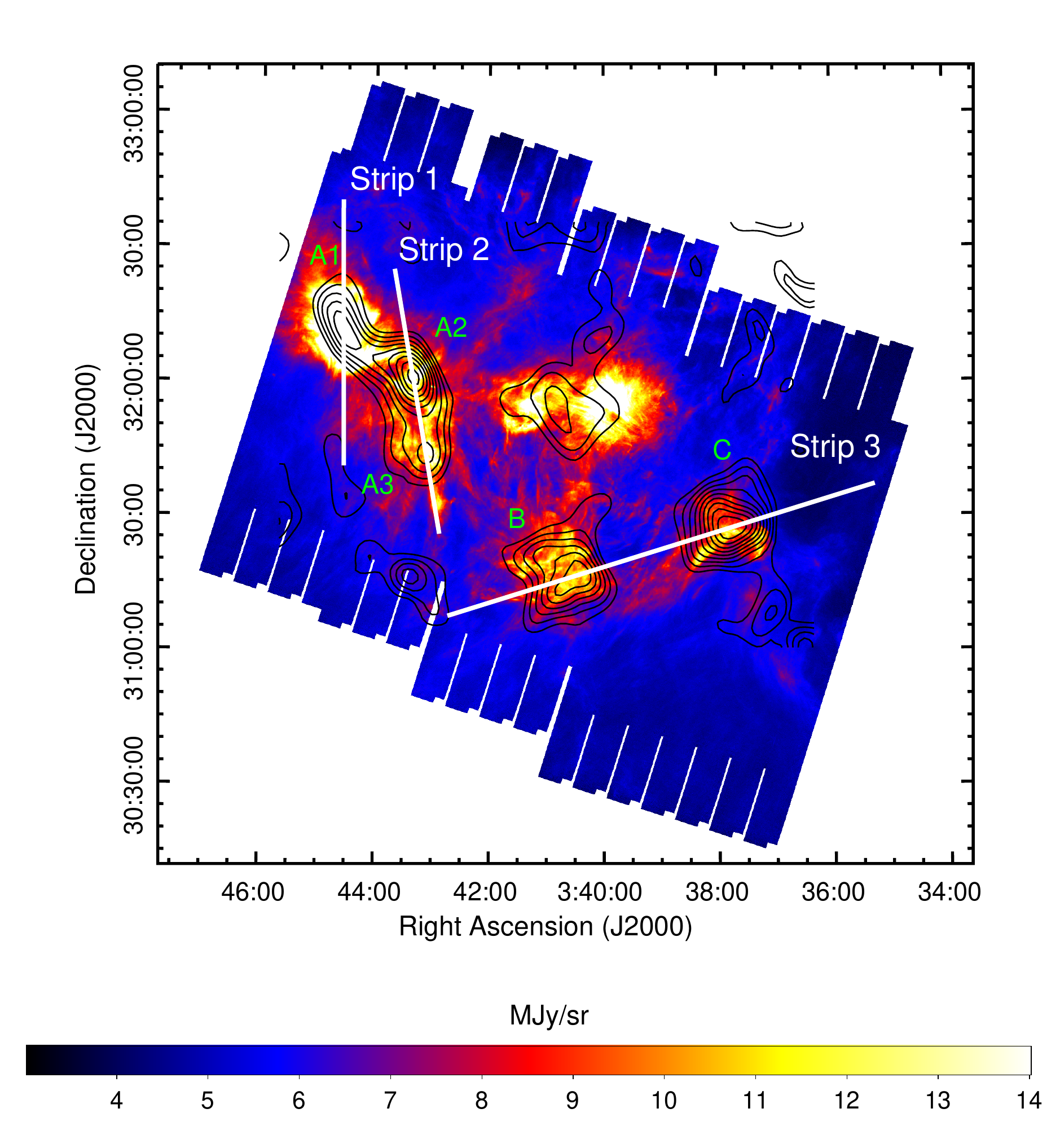} \\ 
 \caption{MIPS 24~$\mu$m image~\citep{Tibbs:11} overlaid with the VSA contours and the location of the three strips~(Strip 1, Strip 2, and Strip 3) observed with the GBT illustrating the coverage of the GBT observations with respect to the AME features~(A1, A2, A3, B, and C) observed with the VSA. The contours correspond to 10, 20, 30, 40, 50, 60, 70, 80, and 90~\% of the peak VSA intensity, which is 200~mJy~beam$^{-1}$.}
\label{Fig:MIPS24_VSA_GBT}
\end{center}
\end{figure}

Given the size of the AME emitting region in the Perseus molecular cloud~\citep[$\sim$~2~degrees $\times$ 2~degrees;][]{Watson:05}, it was not feasible to observe the entire region with the GBT. Therefore, we decided to observe three strips~(Strip 1, Strip 2, and Strip 3) across the region. These strips, displayed in Figure~\ref{Fig:MIPS24_VSA_GBT}, were chosen to coincide with the five AME features~(A1, A2, A3, B, and C) observed with the VSA~\citep{Tibbs:10}, while simultaneously providing enough off-source observations to allow for accurate baseline fitting. We observed the three strips with both the GBT L-Band~(1.4~GHz) and C-Band~(5~GHz) receivers during three days in June 2009 for a total observing time of 14~hrs. Including overheads and calibration observations, the observing time was split with $\sim$~5~hrs for L-Band and~$\sim$~9~hrs for C-Band.

The observations were performed using the Digital Continuum Receiver~(DCR) and the scanning was performed in On-The-Fly mapping mode with a sampling rate of 5~Hz. On-The-Fly mapping involves slewing the telescope across the sky and is the standard method for mapping, or in this case, simply scanning along a single strip, for the GBT. Each of the strips was observed multiple times to increase the total integration time and decrease the noise. Full details of the GBT set-up and observations are listed in Tables~\ref{Table:Summary_GBT_Obs} and~\ref{Table:Summary_GBT_Scans}.

During the observations, a noise diode was repeatedly switched on and off to inject a known level of noise into the system. This was used to convert the raw data to antenna temperature, $T_{ant}$, using

\begin{equation}
T_{ant} = \left \langle \frac{T_{cal}}{P_{cal_{on}} - P_{cal_{off}}} \right \rangle \cdot \frac{(P_{cal_{on}} + P_{cal_{off}})}{2}~\mathrm{K}
\label{equ:gbt_cal}
\end{equation}

\noindent
from~\citet{Maddalena:02}, where $T_{cal}$ is the equivalent temperature of the noise diode in K and $P_{cal_{on}}$ and $P_{cal_{off}}$ are the data observed with the noise diode being switched on and off, respectively.

\begin{figure}
\begin{center}
\includegraphics[angle=0,scale=0.48]{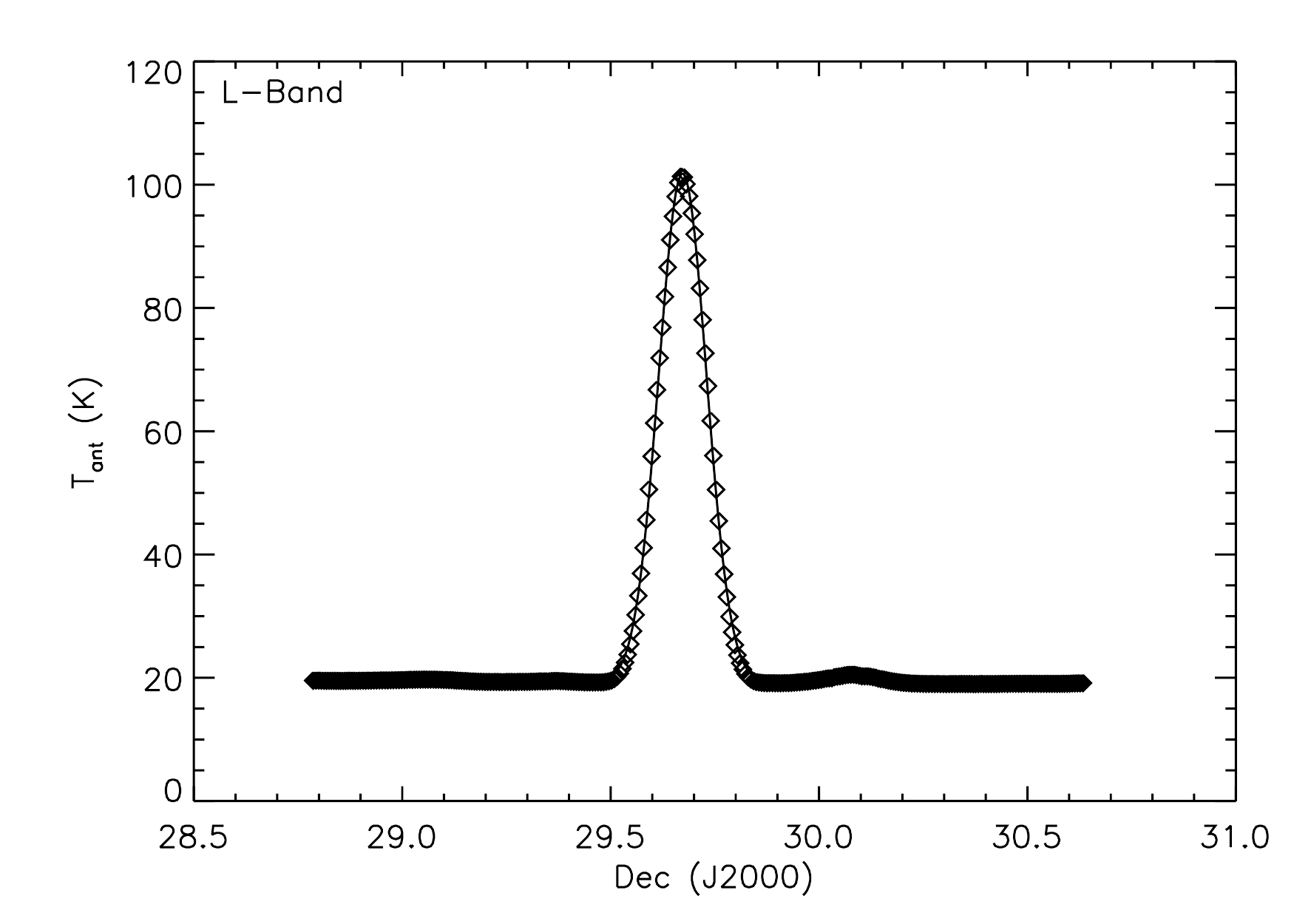} \\
\includegraphics[angle=0,scale=0.48]{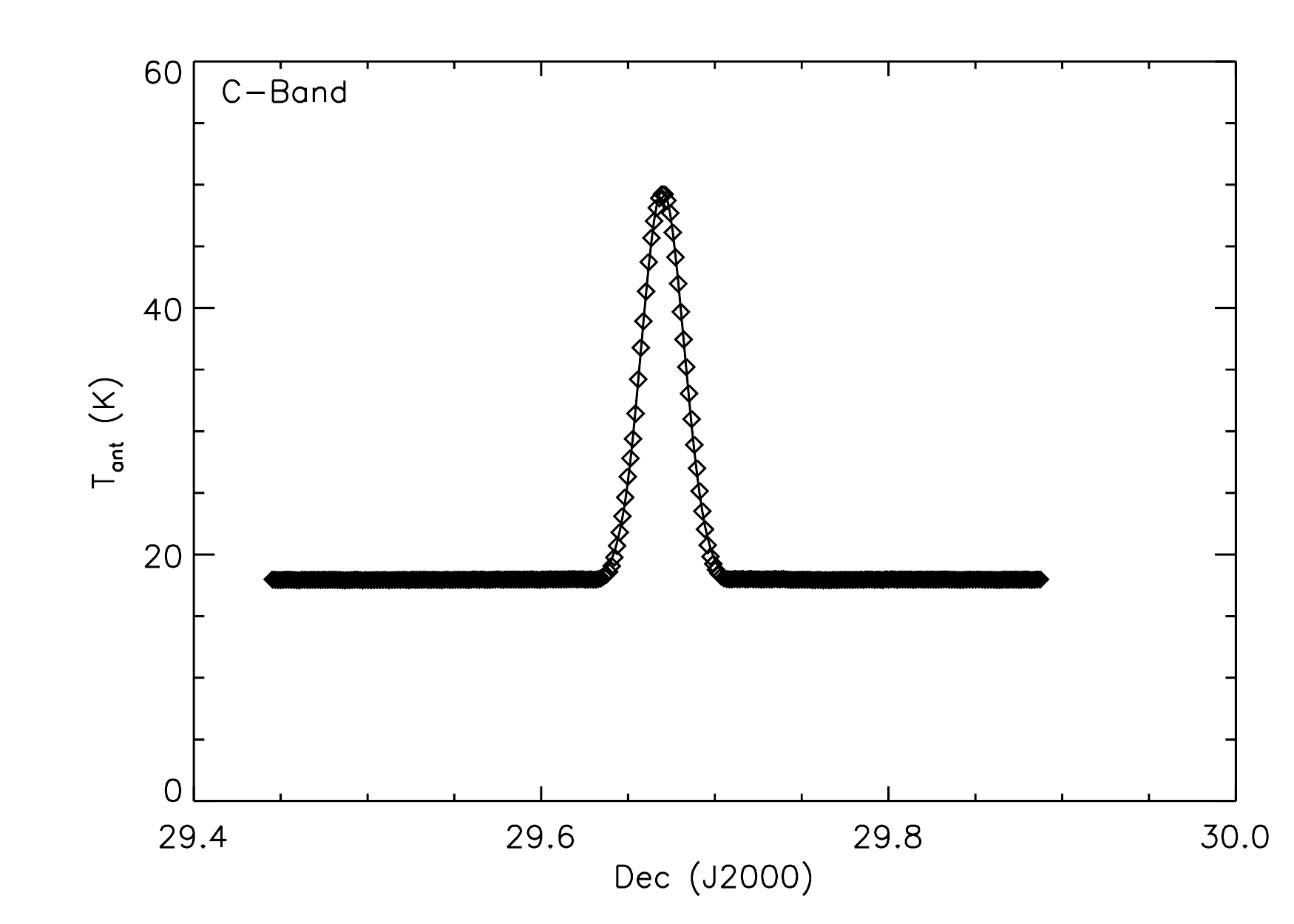} \\ 
\caption{Scans of the calibration source 3C123 at L-Band (top) and C-Band (bottom). The data (open diamonds) were fitted with a Gaussian with a baseline offset (solid line). The observations of 3C123 were used to calibrate the data and convert from an antenna temperature scale to a flux density scale~(see Section~\ref{sec:obs} for details).}
\label{Fig:Peak_Scan_3C123}
\end{center}
\end{figure}

The GBT 1.4 and 5~GHz receiver systems both have two linear polarizations per beam~(XX and YY), which we combined to produce the total power for each band. After converting the data to antenna temperature and combining the two polarizations, the data were converted into flux density units. To do this we used our observations of the calibration source 3C123 that were interspersed with the target observations. These observations involved scanning across 3C123 and were also used to optimize the telescope pointing and focus. We fitted a Gaussian and baseline offset to the observations of 3C123 to obtain antenna temperatures of $T_{ant}$~=~82.45~$\pm$~0.14~K and $T_{ant}$~=~31.61~$\pm$~0.22~K at 1.4 and 5~GHz, respectively. Figure~\ref{Fig:Peak_Scan_3C123} displays one of the scans of 3C123 at both 1.4 and 5~GHz, and the corresponding fit to the data.

\begin{table}
\begin{center}
\caption{GBT Specification for the L-band and C-band Observations}
\begin{tabular}{l c c}
\tableline
\tableline
Parameter & L-Band & C-Band \\ 
\tableline
Receiver & Gregorian L-Band & Gregorian C-Band \\
Back End & DCR & DCR \\
Observing Mode & On-The-Fly & On-The-Fly \\
Central Frequency & 1.4~GHz & 5.0~GHz \\
Bandwidth & 650~MHz & 2000~MHz \\
Beam (FWHM) & 9~arcmin & 2.5~arcmin \\
Scan Speed & 2~arcmin~s$^{-1}$ & 1~arcmin~s$^{-1}$ \\
Sampling Rate & 5~Hz & 5~Hz \\
Theoretical Noise & $\approx$~0.5~mJy~s$^{1/2}$ & $\approx$~0.3~mJy~s$^{1/2}$ \\
R.M.S. Confusion Level & $\approx$~20~mJy & $\approx$~0.7~mJy \\
\tableline
\end{tabular}
\label{Table:Summary_GBT_Obs}
\end{center}
\end{table}

\begin{table*}
\begin{center}
\caption{Summary of the Targeted Strips}
\begin{tabular}{c c c c c c c c c}
\tableline
\tableline
Target & Central RA & Central Dec & Position Angle & Scan Length & \multicolumn{2}{c}{Number of Scans\tablenotemark{a}}$^{}$ & \multicolumn{2}{c}{Noise Levels\tablenotemark{b}} \\ 
 & (J2000) & (J2000)  & (degrees) & (degrees) & \multicolumn{2}{c}{} & \multicolumn{2}{c}{(mJy~beam$^{-1}$)} \\
 & & & & & L-Band & C-Band & L-Band & C-Band \\
\tableline
Strip 1 & 03:44:33.2 & +32:10:59.5 & 180.0 & 0.93 & 24 (90) & 58 (81) & 23.9 (0.31) & 2.7 (0.044) \\
Strip 2 & 03:43:16.7 & +31:55:25.6 &189.8 & 0.93 & 35 (90) & 34 (75) & 12.5 (0.24) & 2.4 (0.057) \\
Strip 3 & 03:38:59.3 & +31:22:10.6 & 287.5 & 1.55 & 20 (75) & 47 (90) & 19.5 (0.28) & 3.0 (0.033) \\
\tableline
\end{tabular}
\label{Table:Summary_GBT_Scans}
\\
\vspace{0mm}
\begin{flushleft}
\hspace{15mm} \footnotesize{$^{a}$Listed are the number of scans used in this analysis along with the total number of scans observed in parentheses.}
\\
\hspace{15mm} \footnotesize{$^{b}$Listed are the r.m.s. noise levels along with the thermal noise levels in parentheses.}
\end{flushleft}
\end{center}
\end{table*}

Based on the flux density calibration observations of~\citet{Ott:94}, we adopted a flux density of 48.01 and 15.95~Jy for 3C123 at 1.4 and 5~GHz, respectively. Therefore, combining the measured antenna temperature of 3C123 with the known flux density, we computed a calibration factor of 0.58~$\pm$~0.01~Jy~K$^{-1}$ at 1.4~GHz and 0.50~$\pm$~0.01~Jy~K$^{-1}$ at 5~GHz. These calibration factors were then applied to the data to convert from antenna temperature to flux density.

\begin{figure*}
\begin{center}
\includegraphics[angle=0,scale=0.95]{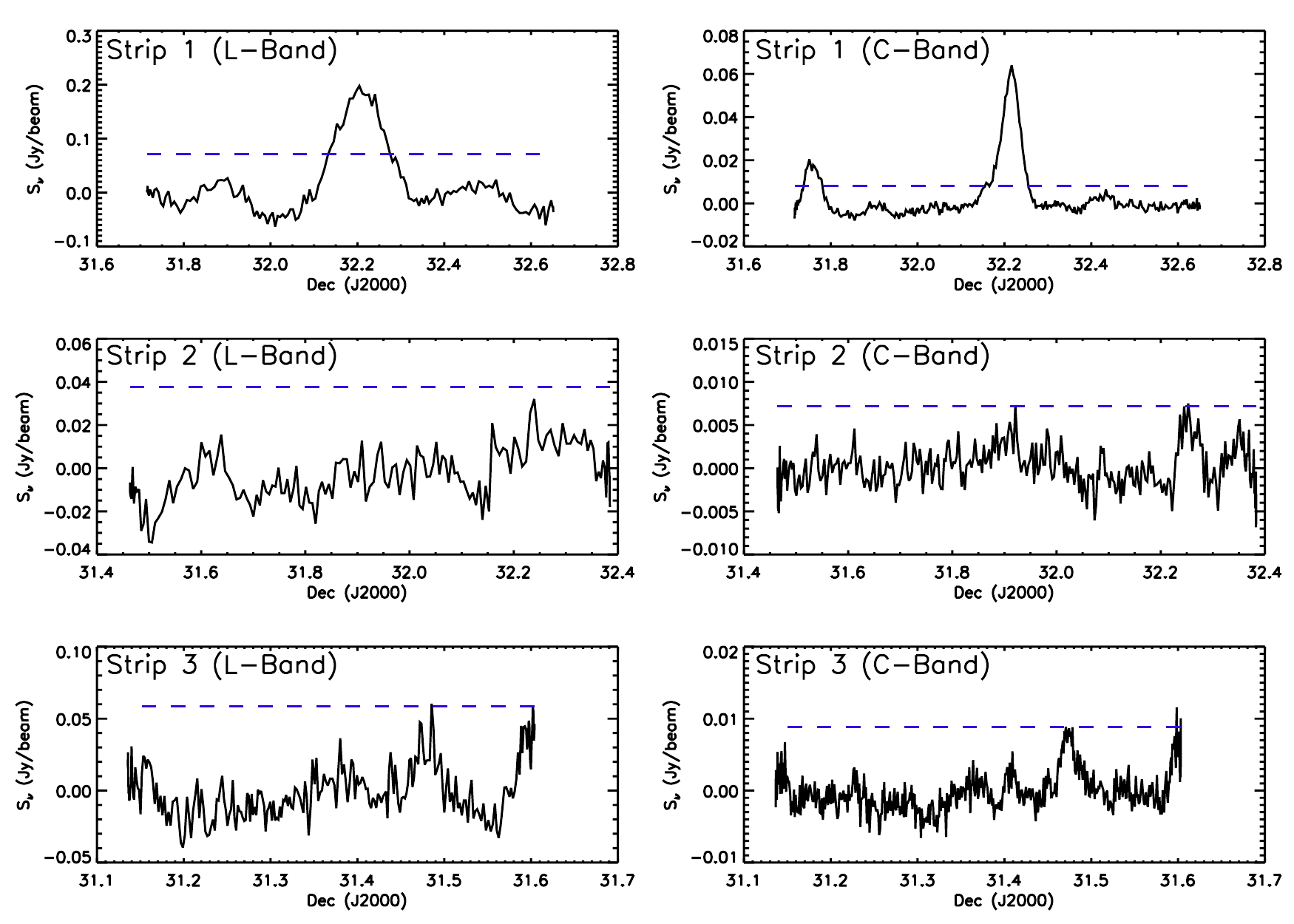} \\ 
 \caption{The final GBT L-Band (left) and C-Band (right) scans for Strips 1~(top), 2~(middle), and 3~(bottom). Also plotted is the 3$\sigma$ upper limit~(dashed line) of the data for each scan. These scans show that we do not detect any significant extended emission. It is also possible to identify the two point sources, NVSS~J034433+321255 and NVSS~J034439+314523, in Strip 1}
\label{Fig:Final_Scans}
\end{center}
\end{figure*}

\begin{figure*}
\begin{center}
\includegraphics[angle=0,scale=0.95]{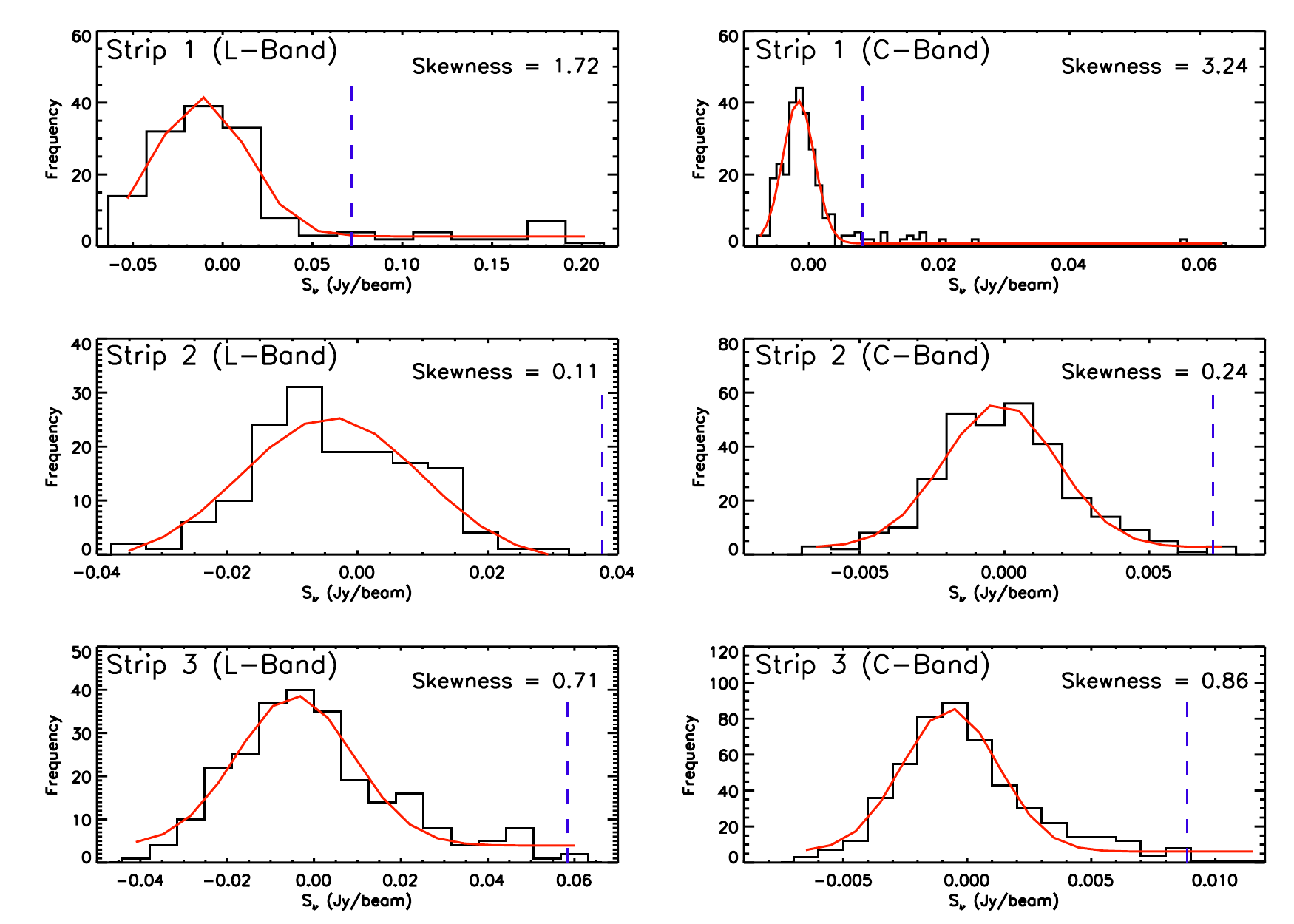} \\ 
 \caption{Histograms of the final GBT L-Band (left) and C-Band (right) scans for Strips 1~(top), 2~(middle), and 3~(bottom). Also displayed is the 3$\sigma$ limit of each distribution~(dashed line) which was computed ignoring the point sources present in Strip 1. By comparing the distributions and the 3$\sigma$ limit it is possible to see that for Strip 1 there is signal present, while for Strips 2 and 3 the data are consistent with noise. Also displayed on the plots is the skewness. Only the L-band and C-band observations of Strip 1 have a skewness $>$ 1, which again suggests that the other strips are dominated by noise.}
\label{Fig:Final_Histograms}
\end{center}
\end{figure*}

\begin{figure}
\begin{center}
\includegraphics[angle=0,scale=0.48]{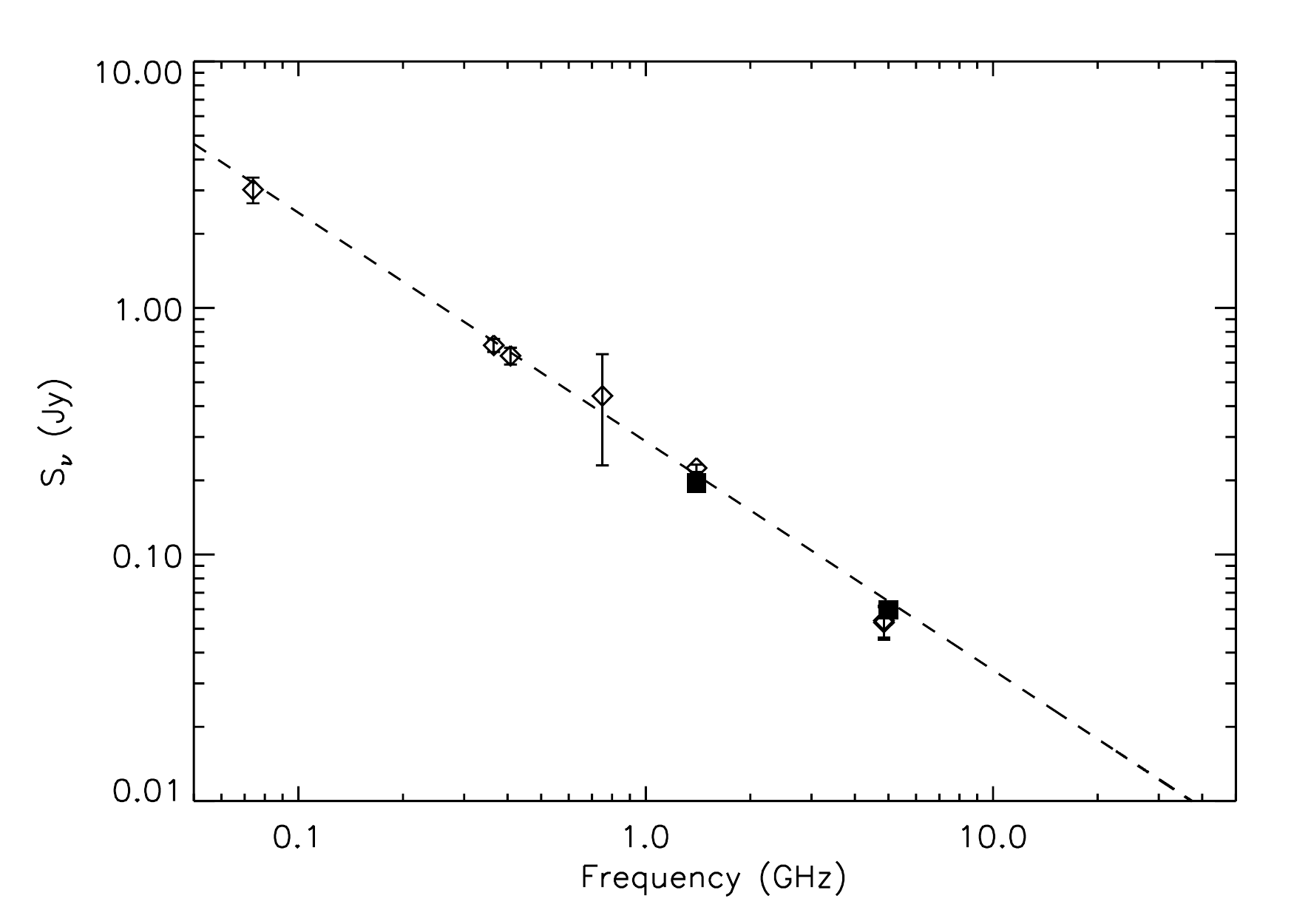} \\ 
 \caption{Spectrum of NVSS J034433+321255. The data from the literature~(open diamonds) have been fitted with a power-law~(dashed line) and the measurements from the GBT data at 1.4 and 5~GHz are overplotted~(filled squares). The consistency between the expected flux densities computed from the fit and the measured values from the GBT data confirm the accuracy of the calibrated data.}
\label{Fig:Source_Cal}
\end{center}
\end{figure}

\begin{figure}
\begin{center}
\includegraphics[angle=0,scale=0.48]{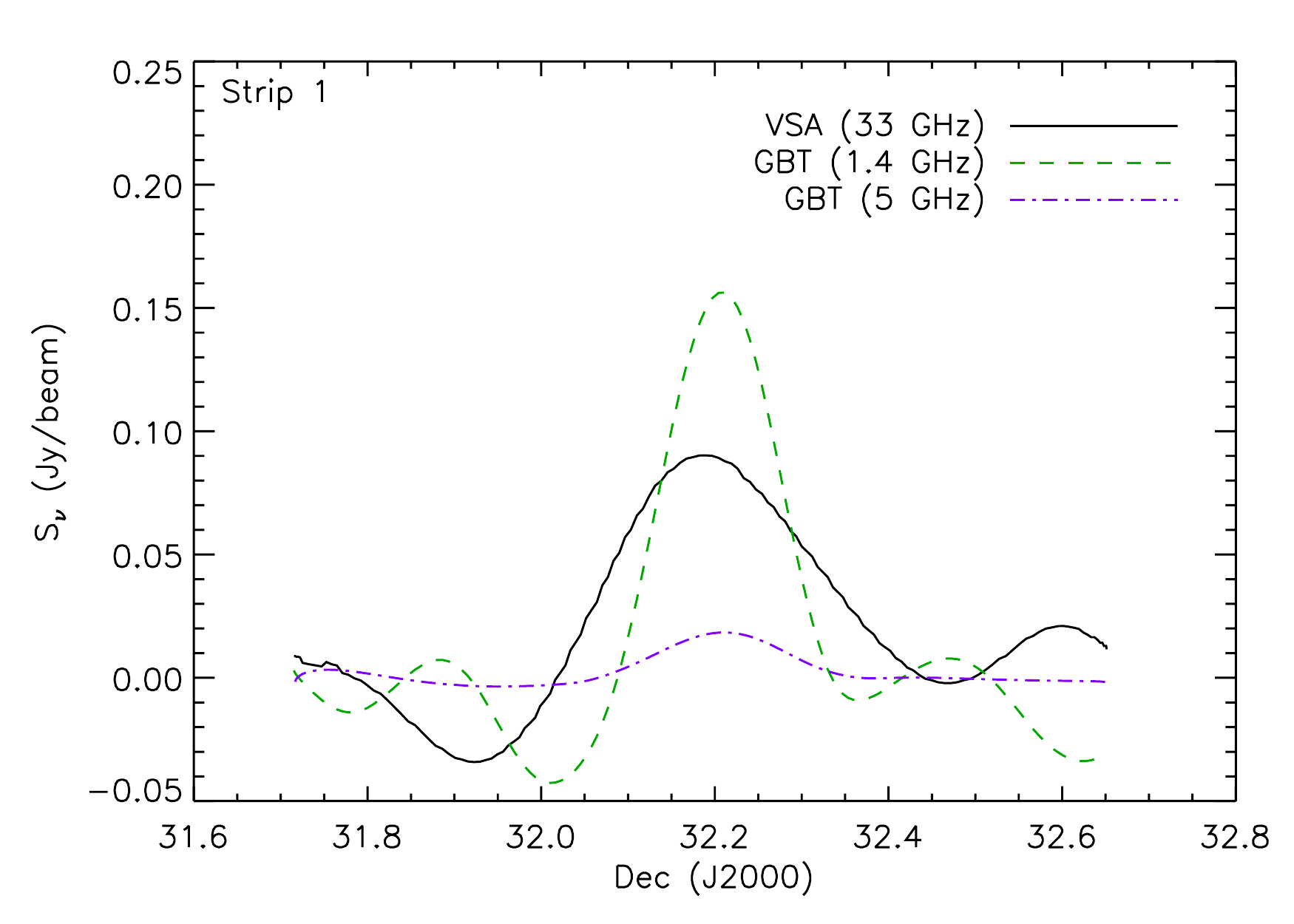} \\ 
\includegraphics[angle=0,scale=0.48]{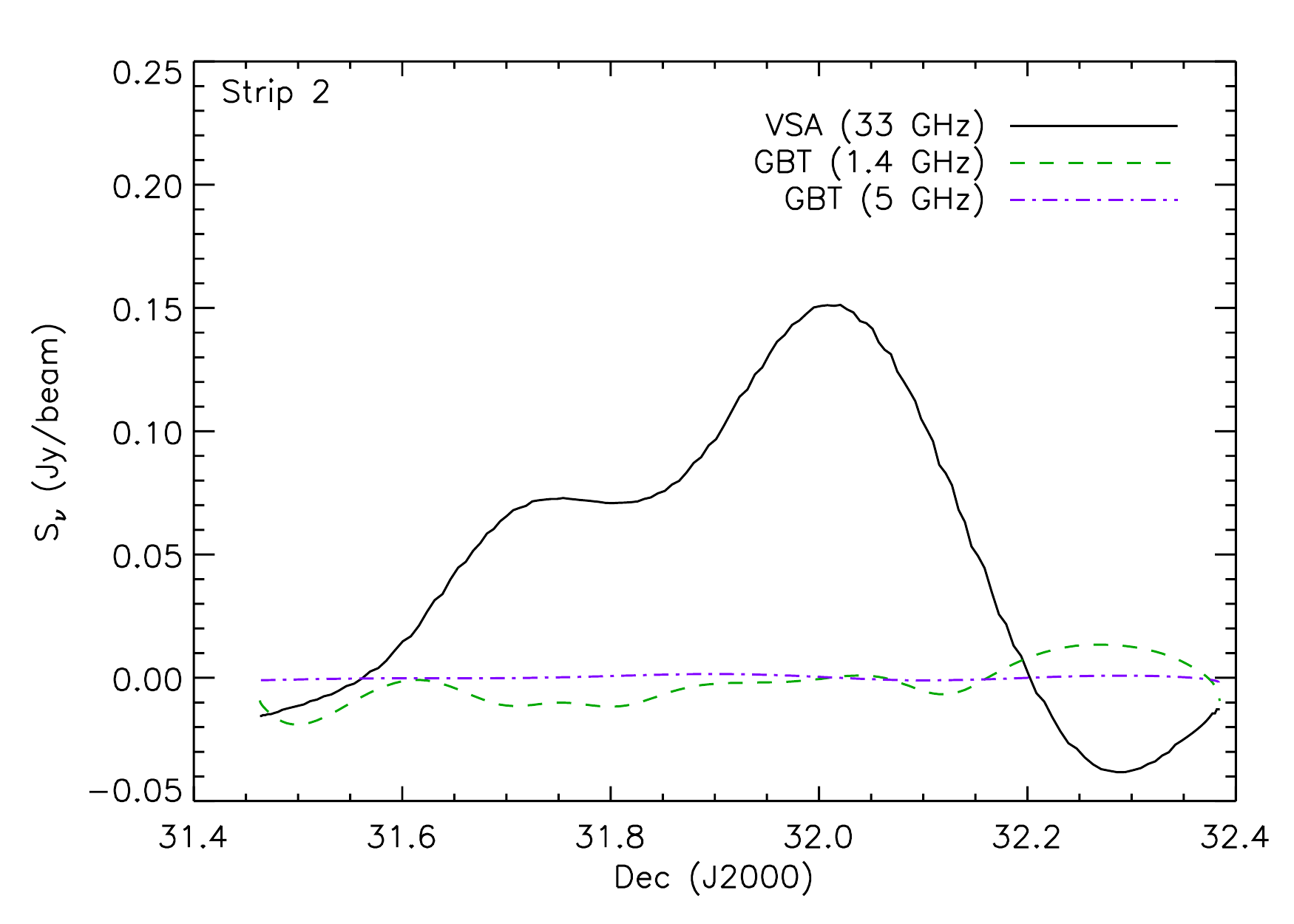} \\ 
\includegraphics[angle=0,scale=0.48]{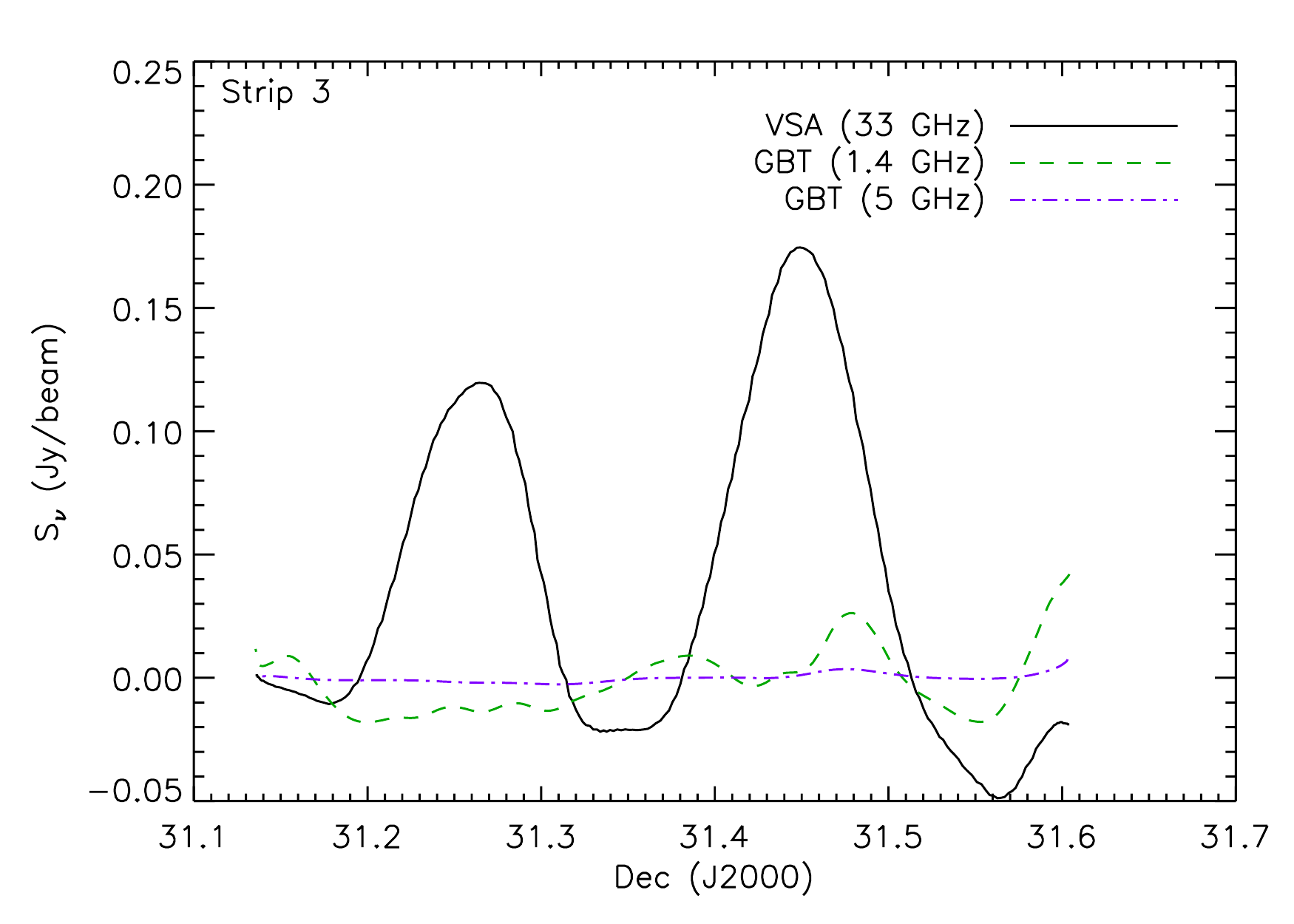} \\ 
 \caption{Comparison of the GBT scans with the VSA observations for Strips 1~(top), 2~(middle), and 3~(bottom). The VSA emission clearly dominates in both Strips 2 and 3, while in Strip 1, the point source NVSS~J034433+321255 appears to dominate. However, when this point source is scaled from 1.4~GHz to 33GHz, as shown in Figure~\ref{Fig:Source_Cal}, the flux density is 11.22~$\pm$~1.11~mJy, which is below the level of the 33~GHz emission. It is also apparent that the spatial structure of the low frequency emission is not comparable to the emission observed at 33~GHz for the three strips.}
\label{Fig:GBT_VSA_Scans}
\end{center}
\end{figure}

\begin{figure}
\begin{center}
\includegraphics[angle=0,scale=0.48]{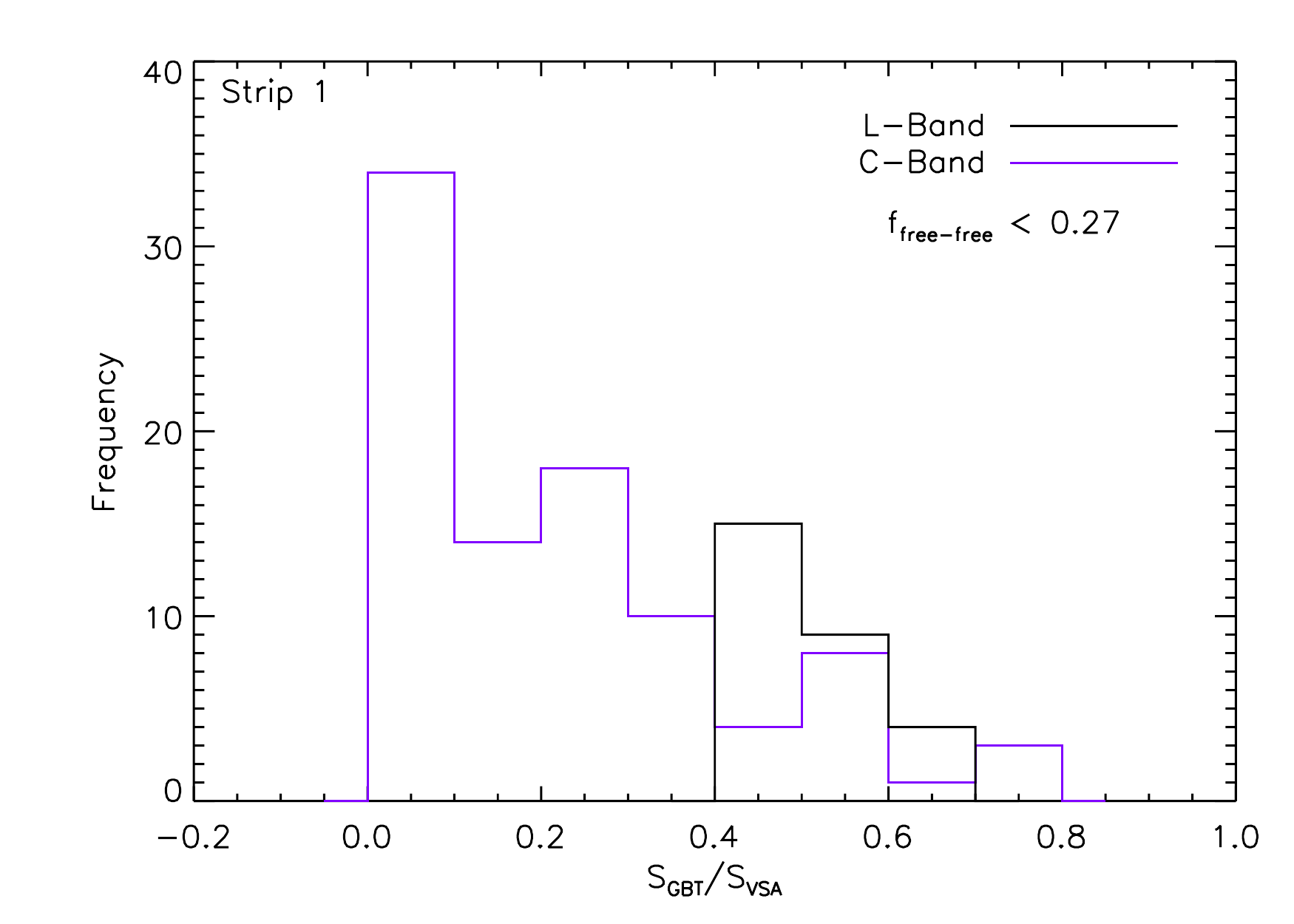} \\ 
\includegraphics[angle=0,scale=0.48]{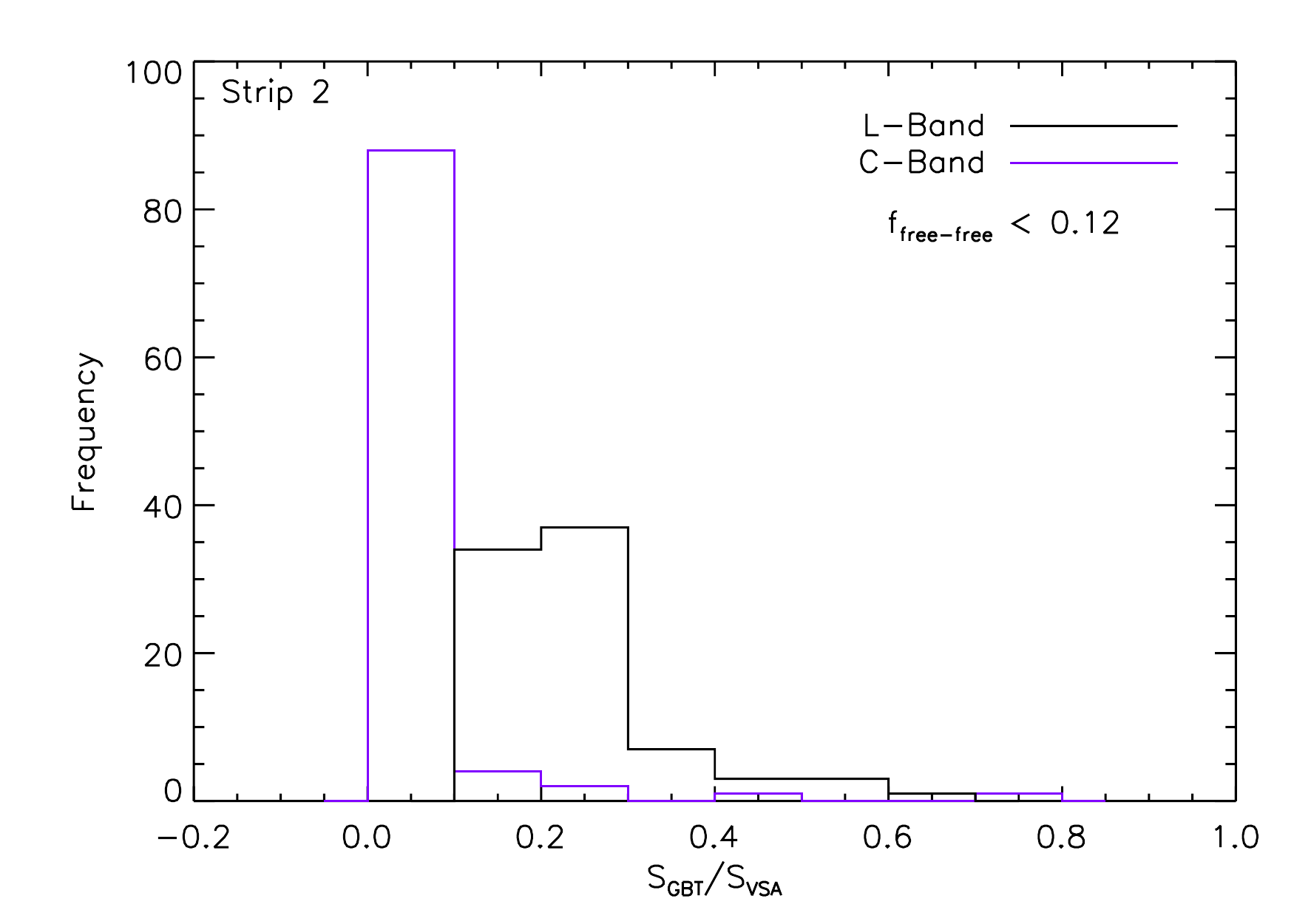} \\ 
\includegraphics[angle=0,scale=0.48]{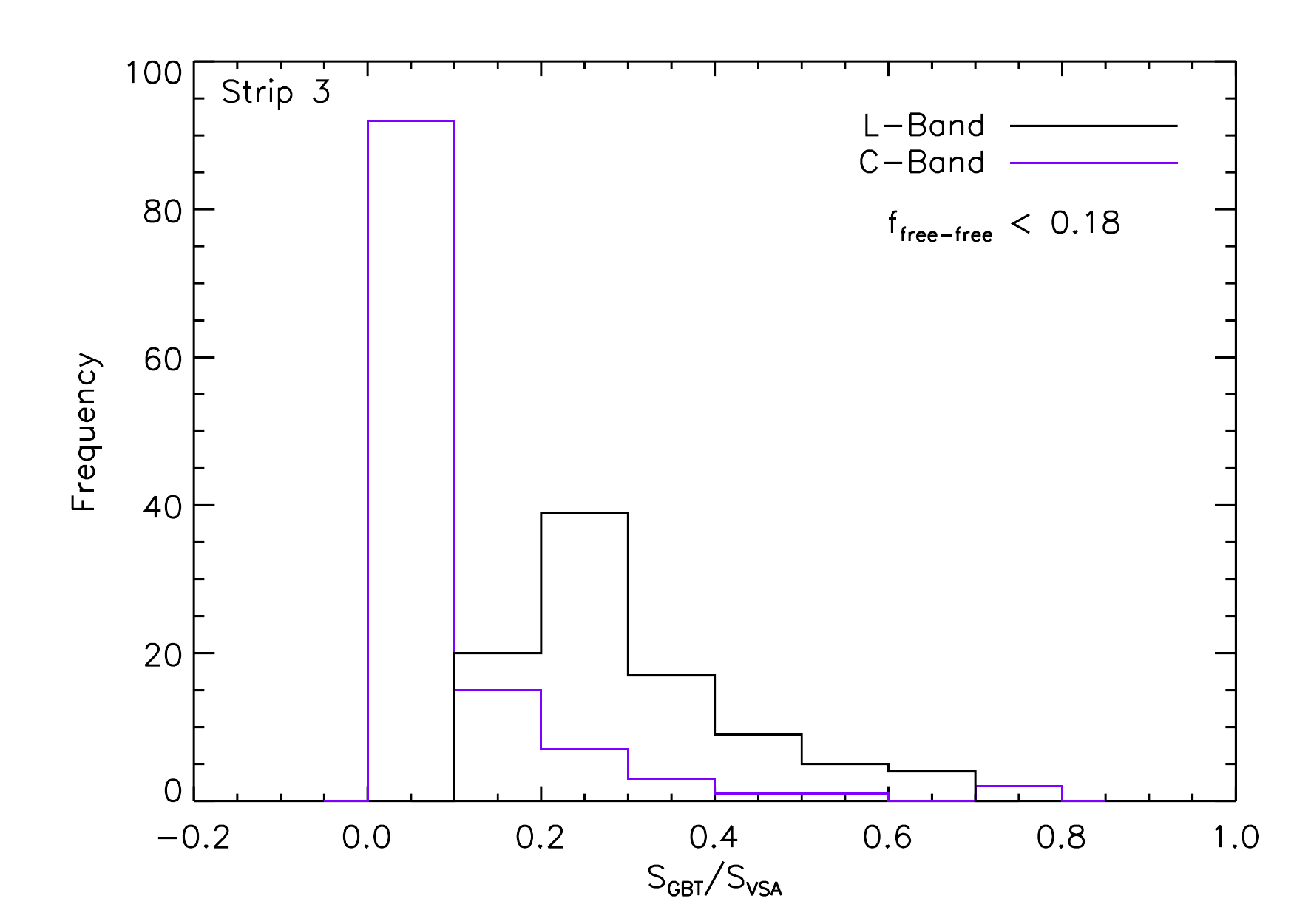} \\ 
 \caption{The distribution of the 3$\sigma$ upper limit of the fraction of free-free emission~(f$_{\mathrm{free-free}}$) at 33~GHz using the L-band and C-Band data for Strips 1~(top), 2~(middle), and 3~(bottom).}
\label{Fig:Fraction_ff}
\end{center}
\end{figure}

Total power observations can be severely affected by the atmospheric opacity. However, at frequencies below 5~GHz, typical values of the zenith opacity are~$\le$~0.01 nepers, which corresponds to an atmospheric attenuation of the order of 1~\% for the elevation of our observations~($\sim$~50~--~80~degrees). However, atmospheric effects are not the only contaminant for total power observations, radio frequency interference~(RFI) and gain variations need to be mitigated. To overcome issues with RFI, all the data were visually inspected and any contaminated data scans were flagged. To help deal with the effects of gain variations, we produced a power spectrum for each data scan. We fitted the power spectrum for the knee frequency, $\nu_{knee}$, above which the data are dominated by white noise only, and the level of the white noise. Since the sampling rate for our GBT observations was 5~Hz, we flagged all data scans for which $\nu_{knee}$~$>$~5~Hz. Finally, to remove the effects of offsets in the data, a baseline subtraction was performed. To determine an accurate baseline level, we binned the data along each scan. The bin sizes were chosen to be approximately equal to the FWHM of the beam, however, we investigated the effects of varying the bin size, and found that the effect was of the order of a few percent. Therefore, we conservatively include a 5~\% uncertainty in the data to include the uncertainty due to the baseline fitting. The median value of the data within each bin was computed, and then the median of all the medians was calculated. This resulted in a median level of the baseline, and we then fitted a first order polynomial to the data within $\pm$~3$\sigma$ of this median value. We only fitted a first order polynomial because a higher order polynomial would potentially remove the structure in which we are interested, while applying the $\pm$~3$\sigma$ cut ensures that any bright sources do not bias the baseline level. The resulting fit was then subtracted from the data. All the data scans for each strip were then combined and the final data scan for each strip was produced by computing the median of the scans. To estimate the thermal noise level in the final data scans, we computed the median of the white noise estimates obtained from fitting the power spectrum, and divided this by the square root of the number of times each strip was observed. For Strips 1, 2, and 3, we estimated a thermal noise level of 0.31, 0.24, and 0.28~mJy~beam$^{-1}$ at L-Band, and 0.044, 0.057, and 0.033 mJy~beam$^{-1}$ for C-Band. These noise levels are consistent with the noise levels obtained by fitting to the power spectrum of the final data scans, confirming that we have been able to reduce the noise level by observing the strips multiple times. However, as we will discuss, the thermal noise is not the dominant source of noise present in the data scans. The final data scans at both L-band and C-band are displayed in Figure~\ref{Fig:Final_Scans} and the final uncertainties on these scans were estimated by combining, in quadrature, a 2~\% uncertainty in the flux density calibration, a 1~\% uncertainty due to the atmospheric opacity, and a 5~\% uncertainty due to the baseline fitting and subtraction. 

In Figure~\ref{Fig:Final_Scans} there are two plots for each strip, one at 1.4~GHz and one at 5~GHz. Looking at these plots it is possible to identify point sources and some extended structures. To determine the significance of these point sources and extended structure we investigated the distribution of the data as shown in the histograms displayed in Figure~\ref{Fig:Final_Histograms}. These histograms show that for all of the scans, the peak of the distribution appears to occur around a flux of 0~Jy~beam$^{-1}$. We computed the skewness for each scan and found that all the scans have a positive skewness, although only the L-Band and C-Band observations of Strip 1 have a skewness $>$ 1, which is a result of the point sources that are clearly present in these scans~(see Figure~\ref{Fig:Final_Scans}). The other strips all have a skewness of $<$ 1 and this, combined with the fact that the distributions peak around 0~Jy~beam$^{-1}$, suggests that these data scans are dominated by noise. It is known that for continuum observations with most GBT receivers, gain fluctuations in the receiver and electronics can considerably degrade the sensitivity, in some cases by more than an order of magnitude~(GBT Proposers Guide July 2012\footnote{https://science.nrao.edu/facilities/gbt/proposing/GBTpg.pdf}). Therefore, to obtain an estimate of the noise, we computed the r.m.s. for each data scan. For Strips 1, 2, and 3 we computed an r.m.s. of 23.9, 12.5, and 19.5~mJy~beam$^{-1}$ for L-Band and 2.7, 2.4, and 3.0~mJy~beam$^{-1}$ for C-band. Note that for Strip 1 we excluded the data corresponding to the point sources from the r.m.s. calculation. It is clear that these noise values are much larger than the thermal noise estimates~(see Table~\ref{Table:Summary_GBT_Scans}) and hence the data scans are not dominated by thermal noise. By comparing the distribution of the data to the 3$\sigma$ limit displayed in Figure~\ref{Fig:Final_Histograms} it is evident that for Strip 1 there is significant signal present, while for Strips 2 and 3 the data are consistent with noise and some non-significant emission. The 3$\sigma$ limit for each scan is also displayed as a dashed horizontal line in Figure~\ref{Fig:Final_Scans}.

In Strip 1, we can see that there is a bright point source at a declination of~$\approx$~32.2~degrees that is visible at both L-Band and C-Band, and there is a less bright point source at a declination of~$\approx$~31.75~degrees that is only visible in C-Band. Both of these point sources are detected at greater than 3$\sigma$. In Strip 2 there is some extended structure present but this is very low-level emission with no emission detected at greater than 3$\sigma$. As in Strip 2, in Strip 3 there is some low-level, non-significant extended structure. There is also a possible hint of a point source at declination~$\approx$~31.48~degrees, however, like the extended structure, this is not a significant detection. Based on a search of the NASA Extragalactic Database\footnote{http://ned.ipac.caltech.edu}~(NED), we believe that the brighter point source in Strip 1 is NVSS~J034433+321255, the weaker point source in Strip 1 is NVSS~J034439+314523, and the hint of a point source in Strip 3 is NVSS~J033727+312808.

\begin{table}
\begin{center}
\caption{Flux densities from the literature for NVSS~J034433+321255}
\begin{tabular}{c c c}
\tableline
\tableline
Frequency & S$_{\nu}$ & Reference \\ 
(GHz) & (mJy) & \\
\tableline
0.074 & 3020~$\pm$~360 & \citet{Cohen:07} \\
0.365 & 706~$\pm$~43.0 & \citet{Douglas:96} \\
0.408 & 640~$\pm$~50.0 & \citet{Colla:70} \\
0.750 & 440~$\pm$~210 & \citet{Pauliny-Toth:66} \\
1.4 & 224.5~$\pm$~7.9 & \citet{Condon:98} \\
4.85 & 53.0~$\pm$~8.0 & \citet{Becker:91} \\ 
4.85 & 54.0~$\pm$~8.0 & \citet{Gregory:91} \\
\tableline
\end{tabular}
\label{Table:NED_Data}
\end{center}
\end{table}

Although we are interested in the extended emission and how it compares to the 33~GHz emission~(see Section~\ref{sec:discuss}), the significant detection of point sources in Strip 1 allows us to check the calibration levels. Therefore, for both point sources in Strip 1, we simultaneously fitted a Gaussian and baseline offset to the data. For NVSS~J034439+314523, we obtained a flux density of 20.17~$\pm$~1.11~mJy in C-Band, although this flux density may be slightly affected by the fact that the source appears at the very edge of the scan. This source is not seen in the L-Band data and this is likely due to a lack of sensitivity. For NVSS~J034433+321255 we observed a flux density of 194.86~$\pm$~10.68~mJy at 1.4~GHz and 59.65~$\pm$~3.29~mJy at 5~GHz. Based on data obtained from NED, which spanned a frequency range from 74~MHz to 4.85~GHz~(see Table~\ref{Table:NED_Data}), we produced a spectrum for NVSS~J034433+321255, which is displayed in Figure~\ref{Fig:Source_Cal}. We fitted the data from the literature with a power-law of the form S$_{\nu}$~$\propto$~$\nu^{\alpha}$ and found a spectral index of $\alpha$~=~$-$0.93~$\pm$~0.01. This fit results in an expected flux density of 210.46~$\pm$~18.02~mJy at 1.4~GHz and 64.61~$\pm$~5.87~mJy at 5~GHz, and these values are consistent with the flux densities measured from the GBT observations. In Figure~\ref{Fig:Source_Cal}, the GBT data at 1.4 and 5~GHz are overplotted on the spectrum and are consistent with the fit to the data from the literature. The consistency between the GBT observations and the values from the literature confirms the accuracy of the GBT data. 



\section{Comparing the GBT Observations with the VSA Observations}
\label{sec:discuss}

Now that we have processed the GBT data and have a measure of the 1.4 and 5~GHz emission in each of the three strips~(Figure~\ref{Fig:Final_Scans}), we wish to compare the level of this low frequency emission with the emission observed at 33~GHz. The 33~GHz emission was observed with the VSA at an angular resolution of~$\sim$~7~arcmin, with a total of 11 individual pointings to cover the entire cloud. A map for each of the pointings was produced using the standard \textsc{aips} routines to perform both the CLEANing and deconvolution. The final map was produced by creating a mosaic of the individual maps, and is sensitive to angular scales of~$\sim$~10~--~40~arcmin~\citep[for further details see][]{Tibbs:10}.

To compare interferometric data and single dish data, the single dish data is usually resampled with the sampling distribution of the interferometer to account for the incomplete sampling in the \textit{u,v} plane. However, since we only have one dimensional GBT scans, this method is not feasible as the result would be completely contaminated by edge effects due to the Fourier transform. Nonetheless, it is still possible to perform a comparison between the GBT and VSA data. Since we fitted a first order polynomial baseline to the GBT scans, the observations are not sensitive to angular scales greater than the length of the strip. As listed in Table~\ref{Table:Summary_GBT_Scans}, the length of these strips is~$\sim$~55~--~90~arcmin and this means that the GBT observations cover the range of angular scales to which the VSA is sensitive. Therefore, comparing the GBT data with the VSA data allows us to determine the extent of the correlation between the emission at 1.4 and 5~GHz and 33~GHz. 

To perform the comparison between the VSA data and the GBT data, we convolved both the C-Band GBT scans and the VSA map~\citep{Tibbs:10} to 9~arcmin to match the angular resolution of the GBT L-Band observations. We then extracted the flux from the convolved VSA map along the three strips observed with the GBT. These data scans were then compared with the GBT data scans, and the results are displayed in Figure~\ref{Fig:GBT_VSA_Scans}. These plots show the 33~GHz emission observed along the three strips with the VSA, over plotted with both the L-Band and C-Band data at 1.4 and 5~GHz, respectively. From Figure~\ref{Fig:GBT_VSA_Scans} it is apparent that the emission at 33~GHz dominates the emission at 1.4 and 5~GHz for Strips 2 and 3. In Strip 1, the point source NVSS~J034433+321255 at 1.4~GHz appears to dominate the VSA emission. However, when the flux density of this source is scaled to 33~GHz, as shown in the spectrum displayed in Figure~\ref{Fig:Source_Cal}, the level of the emission is much less than that observed by the VSA~--~the flux density of the point source at 33~GHz is 11.22~$\pm$~1.11~mJy. There is a possibility that NVSS~J034433+321255 could be a gigahertz peaked source, with a rising spectrum at frequencies greater than 5~GHz. However, looking at Figure~\ref{Fig:GBT_VSA_Scans} it is clear that the spatial structure of the emission at 1.4 and 5~GHz is not similar to the emission observed at 33~GHz for any of the strips. In Strip 1, the VSA detects an extended structure while the GBT only detects the point source NVSS~J034433+321255, implying that even if the point source is a gigahertz peaked source, it is not dominating the VSA emission. Similarly, in Strips 2 and 3, the low-level non-significant emission at 1.4 and 5~GHz does not match the emission observed with the VSA. Therefore, this confirms that the emission observed with the GBT at 1.4 and 5~GHz is much weaker than the emission observed at 33~GHz with the VSA. It should also be noted that that GBT data displayed in Figure~\ref{Fig:GBT_VSA_Scans} have not been scaled to 33~GHz. Assuming a canonical spectral index for free-free emission of~$\alpha$~=~$-$0.12~\citep[e.g.][]{Dickinson:03}, this implies that the expected level of the GBT emission at 33~GHz will be lower than the level plotted in Figure~\ref{Fig:GBT_VSA_Scans}. We also note that the although the GBT observations cover the range of angular scales to which the VSA is sensitive, the emission observed by the VSA on this range of angular scales is not uniformly sampled due to the incomplete sampling of the \textit{u,v} plane. This is not true for the GBT emission, and therefore the VSA flux displayed in Figure~\ref{Fig:GBT_VSA_Scans} can actually be regarded as a lower limit. Additionally, the comparison between the convolved C-Band data and the data extracted from the convolved VSA map is not quite accurate because the convolved C-Band data lacks information in the direction perpendicular to the scan on angular scales greater than its original angular resolution. Therefore, to try and characterize this effect, we performed simulations using the GB6 map of the region~\citep{Condon:89}. We extracted data scans from the GB6 map, convolved them to 9~arcmin and then compared them with the identical data scan extracted from the convolved GB6 map. This comparison allows us to determine the effect of convolving a single scan versus convolving a map, and to perform this comparison we computed the distribution of the ratio of the two simulated scans. We found that this effect is strongly dependent on the position of the selected strip, and based on the location of the three strips in this analysis, the median effect for Strip 1, Strip 2, and Strip 3, was found to be a factor of 1.49, 1.09, and 1.36, respectively. Since the angular resolution of the GB6 data~(3.5~arcmin) and the C-Band data~(2.5~arcmin) are not identical, this effect may be slightly stronger. We note that this issue does not apply to the L-band data.

Therefore, given the fact that we detected no significant extended structure in the three strips, and to account for the issue regarding the convolution of the C-Band data, we conservatively decided to use the 3$\sigma$ upper limits to estimate the fraction of free-free emission at 33~GHz. For each strip we scaled the 3$\sigma$ upper limit to 33~GHz assuming a typical free-free spectral index of $-$0.12 and compared this to the VSA emission. The results of this analysis are plotted as histograms in Figure~\ref{Fig:Fraction_ff}. For each strip, the histogram displays the distribution of the fraction of free-free emission at 33~GHz based on the L-Band and C-Band 3$\sigma$ upper limits. Since we are interested in constraining the fraction of free-free emission in the AME features observed with the VSA at 33~GHz, this analysis was restricted to the regions along each strip in which the 33~GHz emission is greater than the 3$\sigma$ upper limits. Regions in which the 33~GHz emission is less than the 3$\sigma$ upper limits are regions in which the 3$\sigma$ upper limits are an over estimate of the free-free emission. As an estimate of the fraction of free-free emission at 33~GHz, we computed the median of the entire distribution~(both the L-Band and C-Band distributions) for each strip. We find that the conservative 3$\sigma$ upper limit on the fraction of free-free emission at 33~GHz is 27~\%, 12~\%, and 18~\% for Strip 1, Strip 2, and Strip 3, respectively. We tested the robustness of this result by integrating the 33~GHz flux along the scan and comparing it with the integrated 3$\sigma$ upper limits scaled to 33~GHz, and found results consistent with those displayed in Figure~\ref{Fig:Fraction_ff}.

Therefore, from the plots in Figures~\ref{Fig:GBT_VSA_Scans} and~\ref{Fig:Fraction_ff} we conclude that the low frequency emission, extrapolated to 33~GHz, is much fainter than the emission observed at 33~GHz with the VSA. Even if we ignore the C-Band data completely and just use the L-Band data, the 3$\sigma$ upper limit accounts for only 49~\%, 24~\%, and 27~\% of the emission at 33~GHz for Strip 1, Strip 2, and Strip 3, respectively. This confirms that the emission observed with the VSA at 33~GHz is in excess over the free-free emission, and hence is clearly AME. The results of this analysis are in agreement with the analysis performed by~\citet{Tibbs:10} who found that the free-free emission accounted for~$\sim$~20~--~25~\% of the 33~GHz emission. It is also consistent with the analyses performed on much larger angular scales by~\citet{Watson:05} and the~\citet{Planck_Dickinson:11}, who detected the presence of an AME component with a free-free emission fraction of~$\sim$~15~--~20~\% at 33~GHz. These works explained this excess emission as a result of spinning dust emission, which is also consistent with the results of a recent analysis by~\citet{Tibbs:11}, who performed a detailed analysis of the dust properties in this environment, and concluded that the emission could be explained in terms of the spinning dust hypothesis.


\section{Conclusions}
\label{sec:con}

We have used the GBT to observe three strips at 1.4 and 5~GHz that intersect the five regions of AME in the Perseus molecular cloud, which were detected with the VSA~\citep{Tibbs:10}. The data were processed to remove scans affected by RFI and gain variations and the remaining scans were baseline subtracted using a first order polynomial. The scans were then stacked and the median scan was computed. The final data scans were compared with the emission observed in the corresponding strips of the VSA map at 9~arcmin angular resolution, and we found that neither the level of the emission, nor the spatial structure of the emission at 1.4 and 5~GHz, was comparable to the 33~GHz emission. We computed conservative 3$\sigma$ upper limits of the fraction of free-free emission at 33~GHz of 27~\%, 12~\%, and 18~\% for Strip 1, Strip 2, and Strip 3, respectively. Although this analysis is based solely on one dimensional scans, the results are consistent with previous analyses of this region, confirming the low level of free-free emission and the existence of AME in the Perseus molecular cloud.


\section*{Acknowledgments}

We thank A. Noriega-Crespo and S. Carey for stimulating discussions. We also thank the referee for providing detailed comments that have improved the content of this paper. This work has been performed within the framework of a NASA/ADP ROSES-2009 grant, no. 09-ADP09-0059. CD acknowledges support from an STFC Advanced Fellowship and an EU Marie-Curie IRG grant under the FP7. The National Radio Astronomy Observatory is a facility of the National Science Foundation operated under cooperative agreement by Associated Universities, Inc. This research has made use of the NASA/IPAC Extragalactic Database~(NED) which is operated by the Jet Propulsion Laboratory, California Institute of Technology, under contract with the National Aeronautics and Space Administration.




\begin{thebibliography}{}

\bibitem[Ali-Ha{\"i}moud, Hirata \& Dickinson(2009)]{Ali-Haimoud:09} 
Ali-Ha{\"i}moud Y., Hirata C.~M., Dickinson C.,\ 2009, \mnras, 395, 1055 

\bibitem[AMI Consortium: Scaife et al.(2009a)]{Ami:09} 
AMI Consortium: Scaife, A.~M.~M., Hurley-Walker, N., Green, D.~A., et al.\ 2009a, \mnras, 394, L46

\bibitem[AMI Consortium: Scaife et al.(2009b)]{Scaife:09} 
AMI Consortium: Scaife, A.~M.~M., Hurley-Walker, N., Green, D.~A., et al.\ 2009b, \mnras, 400, 1394 

\bibitem[Becker et al.(1991)]{Becker:91} 
Becker, R.~H., White, R.~L., \& Edwards, A.~L.\ 1991, \apjs, 75, 1 

\bibitem[Casassus et al.(2008)]{Casassus:08} 
Casassus, S., Dickinson, C., Cleary, K., et al.\ 2008, \mnras, 391, 1075

\bibitem[Cohen et al.(2007)]{Cohen:07} 
Cohen, A.~S., Lane, W.~M., Cotton, W.~D., et al.\ 2007, \aj, 134, 1245 

\bibitem[Colla et al.(1970)]{Colla:70} 
Colla, G., Fanti, C., Ficarra, A., et al.\ 1970, \aaps, 1, 281 

\bibitem[Condon et al.(1989)]{Condon:89} 
Condon J.~J., Broderick J.~J., Seielstad G.~A.,\ 1989, \aj, 97, 1064 

\bibitem[Condon et al.(1998)]{Condon:98} 
Condon, J.~J., Cotton, W.~D., Greisen, E.~W., et al.\ 1998, \aj, 115, 1693 

\bibitem[Dickinson et al.(2003)]{Dickinson:03} 
Dickinson, C., Davies, R.~D., \& Davis, R.~J.\ 2003, \mnras, 341, 369 

\bibitem[Dickinson et al.(2010)]{Dickinson:10} 
Dickinson, C., Casassus, S., Davies, R.~D., et al.\ 2010, \mnras, 407, 2223 

\bibitem[Douglas et al.(1996)]{Douglas:96} 
Douglas, J.~N., Bash, F.~N., Bozyan, F.~A., Torrence, G.~W., \& Wolfe, C.\ 1996, \aj, 111, 1945 

\bibitem[Draine \& Lazarian(1998)]{DaL:98} 
Draine B.~T., Lazarian A.,\ 1998, \apj, 508, 157 

\bibitem[Draine \& Lazarian(1999)]{DaL:99} 
Draine, B.~T., \& Lazarian, A.\ 1999, \apj, 512, 740 

\bibitem[Draine \& Hensley(2013)]{Draine:13} 
Draine, B.~T., \& Hensley, B.\ 2013, \apj, 765, 159 

\bibitem[Ghosh et al.(2012)]{Ghosh:12} 
Ghosh, T., Banday, A.~J., Jaffe, T., et al.\ 2012, \mnras, 422, 3617 

\bibitem[Gregory \& Condon(1991)]{Gregory:91} 
Gregory, P.~C., \& Condon, J.~J.\ 1991, \apjs, 75, 1011 

\bibitem[Hoang et al.(2010)]{Hoang:10} 
Hoang, T., Draine, B.~T., \& Lazarian, A.\ 2010, \apj, 715, 1462

\bibitem[Hoang et al.(2011)]{Hoang:11} 
Hoang, T., Lazarian, A., \& Draine, B.~T.\ 2011, \apj, 741, 87 

\bibitem[Maddalena(2002)]{Maddalena:02} 
Maddalena, R.~J.\ 2002, Single-Dish Radio Astronomy: Techniques and Applications, 278, 329 

\bibitem[Ott et al.(1994)]{Ott:94} 
Ott, M., Witzel, A., Quirrenbach, A., et al.\ 1994, \aap, 284, 331

\bibitem[Pauliny-Toth et al.(1966)]{Pauliny-Toth:66} 
Pauliny-Toth, I.~I.~K., Wade, C.~M., \& Heeschen, D.~S.\ 1966, \apjs, 13, 65 

\bibitem[Peel et al.(2012)]{Peel:12} 
Peel, M.~W., Dickinson, C., Davies, R.~D., et al.\ 2012, \mnras, 424, 2676 

\bibitem[Planck Collaboration(2011)]{Planck_Dickinson:11} 
Planck Collaboration, Ade, P.~A.~R., Aghanim, N., Arnaud, M., et al.\ 2011, \aap, 536, A20 

\bibitem[Silsbee et al.(2011)]{Silsbee:11} 
Silsbee, K., Ali-Ha{\"i}moud, Y., \& Hirata, C.~M.\ 2011, \mnras, 411, 2750 

\bibitem[Tibbs et al.(2010)]{Tibbs:10} 
Tibbs C.~T., Watson R.~A., Dickinson C., et al.\ 2010, \mnras, 402, 1969 

\bibitem[Tibbs et al.(2011)]{Tibbs:11} 
Tibbs, C.~T., Flagey, N., Paladini, R., et al.\ 2011, \mnras, 418, 1889 

\bibitem[Tibbs et al.(2012)]{Tibbs:12} 
Tibbs, C.~T., Paladini, R., Compiegne, M., et al.\ 2012, \apj, 754, 94 

\bibitem[Tibbs et al.(2013)]{Tibbs:13} 
Tibbs, C.~T., Scaife, A.~M.~M., Dickinson, C., et al.\ 2013, \apj, in press. (arXiv:1303.5501)

\bibitem[Watson et al.(2005)]{Watson:05} 
Watson, R.~A., Rebolo, R., Rubi{\~n}o-Mart{\'{\i}}n, J.~A., et al.\ 2005, \apjl, 624, L89 

\bibitem[Ysard \& Verstraete(2010)]{Ysard:10} 
Ysard, N., \& Verstraete, L.\ 2010, \aap, 509, A12 

\end{thebibliography}
\end{document}